\documentclass[12pt]{article}
\usepackage{amssymb,amsmath}

\newcommand{\Fc}{\mathcal{F}}

\begin{document}

\title{\textbf{Cosmological models with Yang--Mills fields}}

\author{E.~Elizalde$^{1}$\footnote{E-mail addresses:
elizalde@ieec.uab.es, elizalde@math.mit.edu}, A.J.~L\'opez-Revelles$^{1}$\footnote{E-mail
address: alopez@ieec.uab.es}, S.D.~Odintsov$^{1,2,3}$\footnote{E-mail address:
odintsov@ieec.uab.es}, and  S.Yu.~Vernov$^{1,4}$\footnote{E-mail addresses:
vernov@ieec.uab.es, svernov@theory.sinp.msu.ru} \vspace*{3mm} \\
\small $^1$Instituto de Ciencias del Espacio (ICE/CSIC) \, and \\
\small  Institut d'Estudis Espacials de Catalunya (IEEC) \\
\small  Campus UAB, Facultat de Ci\`encies, Torre C5-Parell-2a planta, \\
\small  E-08193 Bellaterra (Barcelona), Spain\\
\small  $^2$Instituci\'o Catalana de Recerca i Estudis Avancats (ICREA), Barcelona, Spain\\
\small  $^3$Tomsk State Pedagogical University, Tomsk, Russia\\
\small  $^4$Skobeltsyn Institute of Nuclear Physics, Lomonosov Moscow State University,\\
\small  Leninskie Gory, GSP-1, 119991, Moscow, Russia
}

\date{ }

\maketitle

\begin{abstract}
Cosmological models with an $SU(2)$ Yang--Mills field are studied.
For a specific model with a minimally coupled Yang--Mills Lagrangian, which includes
an arbitrary function of the second-order term and a fourth-order term, a corresponding
reconstruction program is proposed.  It is shown that the model with
minimal coupling has no de Sitter solutions,
for any nontrivial function of the second-order term.
To get de Sitter solutions, a gravitational model with nonminimally
coupled Yang--Mills fields is then investigated. It is shown that the model
with non-minimal coupling has in fact a de Sitter solution, even
in absence of the cosmological constant term.
\end{abstract}



\maketitle

\section{Introduction}

One of the most important recent results in observational
cosmology has been the discovery that the Universe is speeding up,
rather than slowing down. The combined analysis of data coming from Supernovae Ia (SNe
Ia)~\cite{SN1}, from the cosmic microwave background (CMB) radiation~\cite{WMAP,
Komatsu:2010fb}, large scale structure (LSS)~\cite{LSS}, baryon
acoustic oscillations (BAO)~\cite{Eisenstein:2005su}, and weak
lensing~\cite{Jain:2003tba}, gives strong evidence for the
accelerated cosmic expansion. This
acceleration suggests that the present day Universe is dominated
by a smoothly distributed slowly varying cosmic fluid with
negative pressure, the so-called dark energy. Contemporary cosmological
observation data strongly support the conclusion
that the dark energy equation of state (EoS) parameter is very close to minus one.
These observations also confirm that the Universe is, to high approximation,
isotropic at very large scales.

Various scenarios for the late-time accelerated expansion of the
Universe have been proposed (for reviews, see~\cite{Peebles:2002gy, Padmanabhan:2002ji,
Copeland:2006wr, Durrer:2007re, NO-rev,Review-Nojiri-Odintsov}).
The simplest of them is general relativity with a cosmological constant
(for a review see, e.g.,~\cite{Weinberg:1988cp}).
The standard way to obtain an evolving EoS parameter is to add scalar
fields into a cosmological model. The evolution of the Universe is quite well described by
cosmological models with scalar fields, in particular, by quintom cosmological
models, which are two-field models with a phantom scalar field and an ordinary scalar
one. Quintom models are being actively studied at present
time~\cite{Guo2004,AKV2,Lazkoz,Setare,QuintomModels} (see also~\cite{Quinmodrev1} for reviews).
Some other popular models involve modifications of
gravity, as for instance $F(R)$ gravity, with $F(R)$ an (in principle) arbitrary
function of the scalar curvature $R$ (for recent reviews see,
e.g.,~\cite{NO-rev,Review-Nojiri-Odintsov,Book-Capozziello-Faraoni}).

A possible source of both inflation and the
late-time acceleration of the Universe is a nonminimal coupling between the scalar
curvature and the matter Lagrangian~\cite{matter-1, Allemandi:2005qs}
(see also~\cite{Deruelle:2008fs}).
Such a coupling may be applied for the realization of a dynamical
cancelation of the cosmological constant~\cite{DC}.
Criteria for the viability of such theories have been discussed
in~\cite{criteria-1, Faraoni:2007sn, Bertolami:2007vu}. A coupling between a function of
the scalar curvature and the kinetic term of a massless scalar field
was considered~\cite{Nojiri:2007bt}.

It is known that the coupling between the scalar
curvature and the Lagrangian of the electromagnetic field arises in curved
space-time due to one-loop vacuum-polarization effects in
Quantum Electrodynamics (QED)~\cite{Drummond:1979pp}.
A unified inflation and late-time acceleration model, with the
electromagnetic field coupling to a function of
the scalar curvature, has been proposed in~\cite{Bamba:2008ja} by
using the analyzing procedure in the electromagnetic field~\cite{Bamba-mag-2}.
Inflation driven by a vector field has been
discussed in~\cite{Ford:1989me}. A related scenario is known as vector
inflation~\cite{vector-inflation}. The idea of vector inflation is to
use a vector field as the inflaton, the field which gets non-zero background
value during inflation and drives the inflationary dynamics.

Non-Abelian gauge fields are widely used in particle physics and are being actively studied in cosmology~\cite{Galtsov_Volkov,Moniz,Alfaro,Darian,Barrow,BNO,ELR,Maleknejad_Sheikh-Jabbari,Banijamali,Galtsov}.
Note that string compactifications may naturally lead to an effective-scalar--Yang--Mills--Einstein theory (plus higher-order corrections).
Inflationary cosmology and the late-time
accelerated expansion of the Universe in a non-minimal, non-Abelian gauge theory
(the Yang--Mills (YM) theory), in which a non-Abelian gauge (YM) field plays a
significant role, has been considered
in~\cite{BNO}, where the authors show that the appearance of such non-minimal terms in the
early Universe can be compatible with current formulations of the YM theory coming from a
specific choice for the non-minimal function.
Also in~\cite{BNO} the cosmological reconstruction of the YM theory has been discussed and a corresponding algorithm has been proposed. A new variant of the accelerating
cosmology reconstruction program has been developed in~\cite{ELR}.

A remarkable fact is that the SU(2) Yang--Mills  field admits an isotropic and
homogeneous parametrization by a single scalar function. This parametrization,
which is useful for the reconstruction
program~\cite{BNO,ELR}, has been employed to get an inflationary scenario~\cite{Maleknejad_Sheikh-Jabbari}.
As noted in~\cite{Maleknejad_Sheikh-Jabbari}, the standard YM term, minimally coupling
with gravity, does not lead to inflation, and thus one should add new terms in order to get a convenient inflationary scenario.
In~\cite{Maleknejad_Sheikh-Jabbari} an inflationary scenario, in which slow-roll
inflation is driven by a non-Abelian gauge field minimally coupled to
gravity has been proposed. To achieve this, the authors add a fourth-degree term to the YM Lagrangian.

In this paper, we generalize the reconstruction program in the case of
a YM Lagrangian with a fourth-degree term (see action (\ref{a1}) below). We then get the equation which connects the YM field with the Hubble parameter.
This equation does not depend on the form of the arbitrary function $\Fc$. Thus, assuming an explicit form for the Hubble parameter
as a function of time, we can get the YM field and, after this, construct the function $\Fc$, which corresponds to the given
solution. One of the main results obtained here is that the model with minimal coupling has no de Sitter solution, whatever the nontrivial function $\Fc$ be.

As is well known, de Sitter solutions play a very important role in
cosmological models, because both inflation and the late-time Universe
acceleration can be described as a de Sitter solution with
perturbations.
In order to obtain these solutions, we then consider a gravitational model with nonminimally
coupled YM fields. We succeed in deriving a first-order ordinary differential equation for the YM field
which corresponds to the de Sitter solution. This equation includes an arbitrary parameter. Depending on the value of this parameter, it can be solved in quadratures or else numerically.

\section{Gravitational models with the Yang--Mills fields}

Let us consider  a minimal gravitational coupling of the $SU(2)$
Yang--Mills field in the general theory of relativity, which is
described by the following action:
\begin{equation}\label{a1}
 S = \int d^4 x \, \sqrt{-g} \, \left[
\frac{M_P^2}{2} R + \mathcal{F} (Z)+\frac{\tilde{\kappa}}{384}
(\epsilon^{\alpha\beta\lambda\sigma}F^a_{~\alpha\beta}F^a_{~\lambda\sigma})^2-\Lambda
\right],  \end{equation}
where $g$ is the determinant of the metric tensor $g_{\mu\nu}$,
$R$ is the scalar curvature,  $M_P= M_{Pl}/\sqrt{8\pi}$,  the Planck mass
$M_{\mathrm{Pl}} =  1.2 \times 10^{19}$ GeV, and $\tilde{\kappa}$ is a constant.
 The $SU(2)$ YM field $A_{\mu}^{b}$ has the internal symmetry index $a$,
 the field strength tensor being
\begin{equation}\label{F}
F^a_{\alpha \beta}
      = \partial_\alpha A^a_\beta - \partial_\beta A^a_\alpha + f^{abc} A^b_\alpha A^c_\beta.
\end{equation}
The function $\mathcal{F} (Z)$ is an arbitrary function of
$Z = F_{\mu\nu}^{a} F^{a\mu\nu}$ (summation in terms of the index $a$ is
understood), while the numbers
$f^{abc}$ are structure constants and thus completely antisymmetric.
For the $SU(2)$ group,
\begin{equation}
    f^{abc}={}-\tilde{g} [abc],
\end{equation}
where $\tilde{g}$ is a constant and $[abc]$ the Levi--Civita antisymmetric symbol (we use this
notation instead of $\varepsilon^{abc}$ because we reserve the last one for the Levi--Civita antisymmetric tensor).
Roman indices, $a$, $b$, $c$, will run over $1,\, 2,\, 3$, and the Levi--Civita tensor is given by
\begin{equation}\label{levicivita}
\epsilon^{\alpha\beta\lambda\sigma}=\sqrt{-g}g^{\rho_1\alpha}g^{\rho_2\beta}g^{\rho_3\lambda}g^{\rho_4\sigma}
[\rho_1\rho_2\rho_3\rho_4],\quad
[0123]=1.
\end{equation}
Models of this kind, in the case $\tilde{\kappa}=0$, have been considered
in~\cite{ELR}. The case $\mathcal F (Z)=Z$ has been analysed in~\cite{Maleknejad_Sheikh-Jabbari} where an inflationary scenario,
 in which slow-roll
inflation is driven by a non-Abelian gauge field minimally coupled to
gravity, has been proposed.

The equation of motion for the field $A^a_\mu$ is
 \begin{equation}
 \label{a2}
 \begin{split}
   &\partial_\nu \left\{ \sqrt{-g} \left[
\mathcal F\,' \left( Z \right) \, F^{a \, \nu \mu} +
\frac{\tilde{\kappa}}{192} J \varepsilon^{\nu \mu \alpha \beta}
F^a_{\alpha \beta} \right] \right\} -{}\\&{}- \tilde{g} \sqrt{-g} \, [abc] \, A^b_\nu \left\{ \mathcal
F\,' \left( Z \right) \, F^{c \, \nu
\mu} - \frac{\tilde{\kappa}}{192} J \varepsilon^{\mu \nu \alpha \beta} F^c_{\alpha \beta} \right\} = 0,
\end{split}
 \end{equation} where
$J\equiv\epsilon^{\alpha\beta\lambda\sigma}F^b_{~\alpha\beta}F^b_{~\lambda\sigma}$.
      Variation of (\ref{a1}) with respect to $g^{\mu \nu}$ yields the  field equations:
 \begin{equation}\label{a3}
   \frac{M_P^2}{2} \left( R_{\mu \nu} - \frac{1}{2} g_{\mu \nu} R
\right) - \frac{1}{2} g_{\mu \nu} \mathcal F \left( Z \right) + 2
\mathcal F\,' \left( Z \right) \, F^a_{\mu \rho} F^{a \, \rho}_\nu -$$
$$- \frac{\tilde{\kappa}}{384}
\left\{ \frac{3}{2} J^2 g_{\mu\nu} - 8J\sqrt{-g}g^{\rho_2\beta}g^{\rho_3\lambda}g^{\rho_4\sigma}
[\mu\rho_2\rho_3\rho_4]\, F^b_{~\nu\beta}F^b_{~\lambda\sigma} \right\} +
\frac{1}{2} g_{\mu \nu} \Lambda = 0.
\end{equation}
In the spatially flat Friedmann--Lema\^{i}tre--Robertson--Walker (FLRW)
universe
\begin{equation}
\label{mFr} ds^2={}-dt^2+a^2(t)\left(dx_1^2+dx_2^2+dx_3^2\right),
\end{equation}
the following ansatz for the $SU(2)$ field:
\begin{equation}
\label{ansatzA}
 A_\mu^b=\left(0,\phi(t)\delta^b_i\right).
\end{equation}
is very useful~\cite{BNO,ELR,Maleknejad_Sheikh-Jabbari}.  The ansatz
identifies the combination of the Yang--Mills fields for which the
rotation symmetry violation is compensated by the gauge
transformations. Thus, we get the rotationally invariant energy--momentum
tensor of the Yang--Mills fields, namely
\begin{equation}
F^b_{0j}={}-F^b_{j0}=\dot\phi\delta^b_j,\quad F^b_{ij}={}-\tilde{g}\phi^2 [bij],
\quad
F^{b0j}={}-F^{bj0}={}-\frac{\dot\phi}{a^2}\delta^{bj},\quad F^{bij}={}-\frac{\tilde{g}\phi^2}{a^4} [bij]
\end{equation}
and
\begin{equation}
    Z\equiv F_{\mu\nu}^{b}F^{b\mu\nu}=6\left(\frac{\tilde{g}^2\phi^4}{a^4}-\frac{\dot\phi^2}{a^2}\right),
\end{equation}
where differentiation with respect to time $t$ is denoted by a dot.

Use of the ansatz (\ref{ansatzA}) allows to obtain the YM energy--momentum
tensor, having the same form as an ideal isotropic fluid with the
energy density $\rho$ and the pressure $P$, in other words: \
$T^{(\mathrm{YM})\mu}_{\qquad\nu} = \mathrm{diag} (- \rho,P,P,P)$.

Taking into account (\ref{levicivita}) and (\ref{mFr}), the equation of motion (\ref{a2}) can be written as follows:
\begin{equation}
 \label{a2}
   \partial_\nu \left\{ \sqrt{-g}
\mathcal F\,' \left( Z \right) \, F^{a \, \nu \mu} -
\frac{\tilde{\kappa}}{192} J [\nu \mu \alpha \beta]
F^a_{\alpha \beta} \right\} -$$
$$- \tilde{g} \, [abc] \, A^b_\nu \left\{ \sqrt{-g} \mathcal
F\,' \left( Z \right) \, F^{c \, \nu
\mu} + \frac{\tilde{\kappa}}{192} J [\mu \nu \alpha \beta] F^c_{\alpha \beta} \right\} = 0.
 \end{equation}
It is convenient to write the Friedmann equations in terms of $\psi\equiv \phi/a$.
Using
\begin{equation}
\label{psiphi}
\dot\phi=a\left(\dot\psi+H\psi\right),
\qquad
\ddot\phi=a\left(\ddot\psi+2H\dot\psi+\psi\left(\dot H+H^2\right)\right),
\end{equation}
where $H=\dot{a}/a$ is the Hubble
parameter, we get the  equations which follow.

       The $(0,0)$ component of (\ref{a3}) reduces to:
 \begin{equation}\label{a5}    \frac{3 M_P^2}{2} H^2 + \frac{1}{2}
\mathcal F \left( Z \right) + 6 \mathcal F\,' \left( Z \right) \,
\left(\dot\psi+H\psi\right)^2
-\frac{3}{4}\tilde{\kappa}\tilde{g}^2\psi^4(\dot\psi+H\psi)^2 - \frac{1}{2} \Lambda = 0,
 \end{equation} where $\psi\equiv \phi/a$. Note that
\begin{equation}
    Z=6\left(\tilde{g}^2\psi^4-(\dot\psi+H\psi)^2\right),
\quad \dot Z=12\left(2\tilde{g}^2\psi^3\dot\psi-(\dot\psi+H\psi)(\ddot
\psi+H\dot\phi+\dot H\psi)\right),$$
$$ J = 24 \tilde g \psi^2 \left( \dot\psi
+H\psi \right).
\end{equation}
  The $(i,i)$ components of (\ref{a3}) yield:
 \begin{equation}\label{a6}    \frac{1}{2} M_P^2 \left[ 2 \dot H + 3
H^2 \right] + \frac{1}{2} \mathcal F \left( Z \right) + 2 \mathcal F\,' \left( Z
\right) \left[ \left(\dot\psi+H\psi\right)^2 - 2 \tilde{g}^2 \psi^4
\right]-\frac{3}{4}\tilde{\kappa}\tilde{g}^2\psi^4(\dot\psi+H\psi)^2 - \frac{1}{2} \Lambda =
0.  \end{equation}
      By subtracting Eq.~(\ref{a5}) from Eq.~(\ref{a6}), we get
\begin{equation}
\frac{M_P^2}{2} \dot{H}= 2\mathcal F\,' \left( Z \right) \left(
\left(\dot\psi+H\psi\right)^2 + \tilde{g}^2 \psi^4\right).
\end{equation}
From this equation, it follows that the model considered does not have nontrivial de
Sitter solutions ($H$ is a constant). Such solutions can exist only if
either $\mathcal F (Z)$ is a constant, or the function $\psi(t)=0$. In the next section
we show that nontrivial de Sitter solutions do exist in a model which has a
non-minimal coupling.

Rewriting the last equation in the following form
 \begin{equation}\label{a7}    \mathcal F\,' \left( Z \right) =
\frac{M_P^2}{4} \dot{H}  \left( \left(\dot\psi+H\psi\right)^2 +
\tilde{g}^2 \psi^4\right)^{-1},
\end{equation}
introducing (\ref{a7}) into Eq. (\ref{a5}) and differentiating, we
get:
 \begin{equation}\label{a8}
 \begin{split}
H\dot H&+\frac{\dot{H}}{2\left( \left(\dot\psi+H\psi\right)^2 +
\tilde{g}^2 \psi^4\right)}\left(2\tilde{g}^2\psi^3\dot\psi
-\left(\dot\psi+H\psi\right)\left(\ddot \psi+H\dot\psi+\dot H\psi\right)\right)+{}\\
&{}+\frac{1}{2}\left(\dot\psi+H\psi\right)^2\frac{d}{dt}\left[\frac{\dot{H}
}{ \left(\dot\psi+H\psi\right)^2 + \tilde{g}^2
\psi^4}\right]+\frac{\dot{H}
\left(\dot\psi+H\psi\right)\left(\ddot\psi+H\dot\psi+\dot H\psi\right)
}{ \left(\dot\psi+H\psi\right)^2
+ \tilde{g}^2 \psi^4}-{}\\
&{}-
\frac{\tilde{\kappa}}{M_P^2}\tilde{g}^2\left(\psi^3\left(\dot\psi+H\psi\right)^2\dot\psi+
\frac{1}{2}\psi^4\left(\dot\psi+H\psi\right)\left(\ddot\psi+H\dot\psi+\dot
H\psi\right)\right)=0.
\end{split}
 \end{equation}

If we assume, or rather know, the specific form of the Hubble function $H(t)$,
then Eq.~(\ref{a8}) constitutes a differential equation for $\psi(t)$ and, once
we determine this function, the corresponding Yang--Mills theory can be found (i.e., the
function $\mathcal F \left( Z \right)$) which reproduces the cosmology
given by $H(t)$ in the frame of the spatially flat FLRW universe. From
(\ref{a7}) we can find $\mathcal F \left( Z \right)$ up to an
integration constant. This constant can be determined from (\ref{a6})
and corresponds to the cosmological constant.

\section{Non-minimal gravitational coupling with the Yang--Mills field}

\subsection{Action and equations} In this section we will consider  a
non-minimal gravitational coupling of the $SU(2)$ YM field in general
relativity, which is described by the action:
\begin{eqnarray}
S_{\mathrm{GR}} &=&
\int d^{4}x \sqrt{-g}
\left[ \frac{M_P^2}{2} R
+{\mathcal{L}}_{\mathrm{YM}}-\Lambda
\right]\,,
\label{eq:2.1} \\[2mm]
{\mathcal{L}}_{\mathrm{YM}}\!
&=&\!\!{}-\frac{1}{4} (1+f(R)) Z,
\label{eq:2.3}
\end{eqnarray}
where $f(R)$ is an  arbitrary, thrice differentiable function of $R$.

The field equations can be derived by taking variations of the
action in Eq.~(\ref{eq:2.1}) with respect to the
metric $g_{\mu\nu}$ and the $SU(2)$ YM field $A_{\mu}^{a}$, as follows:
\begin{equation}
R_{\mu \nu} - \frac{1}{2}g_{\mu \nu}R
=\frac{1}{M_P^2}\left(T^{(\mathrm{YM})}_{\mu \nu}-\Lambda g_{\mu \nu}\right)\,,
\label{eq:2.9}
\end{equation}
with
\begin{eqnarray}
T^{(\mathrm{YM})}_{\mu \nu}
 &=&
\left(1+f(R)\right) \left( g^{\alpha\beta} F_{\mu\beta}^{b} F_{\nu\alpha}^{b}
-\frac{1}{4} g_{\mu\nu} \mathcal{F} \right)+{}
\nonumber \\[2mm]
&+&\frac{1}{2} \left\{ f^{\prime}(R) \mathcal{F} R_{\mu \nu}
+ g_{\mu \nu} \Box \left[ f^{\prime}(R) \mathcal{F} \right]
- {\nabla}_{\mu} {\partial}_{\nu}
\left[ f^{\prime}(R) \mathcal{F} \right]
\right\},
\label{eq:2.10}
\end{eqnarray}
where the prime denotes derivative with respect to $R$,
${\nabla}_{\mu}$ is the covariant derivative operator associated with
$g_{\mu \nu}$, and $\Box \equiv g^{\mu \nu} {\nabla}_{\mu} {\partial}_{\nu}$
is the covariant d'Alembertian for the scalar field.

It is convenient to write down the trace equation
\begin{equation}
R={}-\frac{1}{M_P^2}g^{\mu\nu}\left(T^{(\mathrm{YM})}_{\mu
\nu}-g_{\mu\nu}\Lambda\right)={}-\frac{1}{2M_P^2} \left\{
f^{\prime}(R) \mathcal{F} R + 3 \Box \left[ f^{\prime}(R)
\mathcal{F} \right]-8\Lambda\right\}\,. \label{trace_eq}
\end{equation}
We will show that the trace equation is useful in order to find the de Sitter solutions.

\subsection{The Friedmann--Lema\^{i}tre--Robertson--Walker metric and
equations of motion}

Using the ansatz~(\ref{ansatzA}), we get the following equations in the FLRW metric
(see the Appendix, for details):
\begin{equation}
\begin{split}
&3H^2=\frac{1}{M_P^2}(\Lambda+\rho)=\frac{\Lambda}{M_P^2}+\frac{3}{2M_P^2}\left[(1+f(R))
\left(\tilde{g}^2\psi^4+(\dot\psi+H\psi)^2\right)-{}\right.\\
&\left.{}- 6\left( \dot{H} + H^2 \right)f^{\prime}(R) \left(
\tilde{g}^2\psi^4-(\dot\psi+H\psi)^2\right)
+ 6H\partial_0\left[f^{\prime}(R)\left(\tilde{g}^2\psi^4- \left(\dot\psi+H\psi\right)^2\right)\right]\right],
\label{equa00psi}
\end{split}
\end{equation}
\begin{equation}
\begin{split}
\label{equajjpsi}
    2\dot H +3H^2=\frac{\Lambda-P}{M_P^2}=&\frac{\Lambda}{M_P^2}-\frac{1}{2M_P^2}\left[(1+f(R))
    \left(\tilde{g}^2\psi^4+\left(\dot\psi+H\psi\right)^2\right)
+{}\right.\\+ &\, 6\left(\dot
H+3H^2\right)f'(R)\left(\tilde{g}^2\psi^4-\left(\dot\psi+H\psi\right)^2\right)
    -{}\\- & \left.6[\partial_0\partial_0+2H\partial_0]
     f^{\prime}(R) \left(\tilde{g}^2\psi^4-\left(\dot\psi+H\psi\right)^2\right)\right],
\end{split}
\end{equation}
It is suitable to get, from system (\ref{equa00psi})--(\ref{equajjpsi}), the following equivalent one:
\begin{equation}
\label{equaRLambda}
    R=\frac{1}{M_P^2}\left(4\Lambda-3R\vartheta
    +9\ddot\vartheta+27H\dot\vartheta \right),
\end{equation}
\begin{equation}
\label{equa_dotH} \dot H={}-\frac{1}{2M_P^2}\left[2(1+f(R))
\left(\tilde{g}^2\psi^4+(\dot\psi+H\psi)^2\right)-6\dot H
\vartheta-3\ddot\vartheta+3H\dot\vartheta\right],
\end{equation}
where
\begin{equation}
 \vartheta\equiv f^{\prime}(R) \left(\tilde{g}^2\psi^4-(\dot\psi+H\psi)^2\right).
\end{equation}
We can see that the term
$(1+f(R))\left(\tilde{g}^2\psi^4+(\dot\psi+H\psi)^2\right)$ corresponds to
radiation since, if we  neglect other terms, we get $\rho=3P$. This
result is a trivial generalization of the corresponding one in the
model with minimal coupling ($f(R)=0$), considered
in~\cite{Maleknejad_Sheikh-Jabbari}. In the $f(R)$ modified model,
$T^{(\mathrm{YM})}_{\mu\nu}$ has also terms proportional to $f'(R)$. In
this paper, we will show that these terms can actually play the role of the
cosmological constant.

\subsection{Solutions with constant Hubble parameter}

We now investigate the de Sitter solutions for the model (\ref{eq:2.1}).
Our goal is to see how the YM field, which is described by
${\mathcal{L}}_{\mathrm{YM}}$, can change the value of the cosmological
constant. In particular, we will demonstrate in this section, that there
do exist de Sitter solutions in the case when $\Lambda=0$.

We seek solutions with $H=H_0=$const, in other words, de Sitter and Minkowski solutions.
If $H=H_0$, then $R=R_0=12H_0^2$, and (\ref{equaRLambda}) is a linear differential equation in $\vartheta$:
\begin{equation}
\label{equa_vartheta}
        \ddot\vartheta+3H_0\dot\vartheta-4H_0^2\vartheta=B,
\end{equation}
where the constant $B=(M_P^2R_0-4\Lambda)/3$. Eq.~(\ref{equa_vartheta})
has the following general solution
\begin{equation}
\label{dSsolequ1}
\vartheta=C_1e^{Ht}+C_2e^{-4Ht}-\frac{B}{4H^2}
\end{equation}
and, from (\ref{equa_dotH}), we get
\begin{equation}
\label{equa_dotH2}
2(1+f(R_0))\left(\tilde{g}^2\psi^4+(\dot\psi+H\psi)^2\right)-3\ddot\vartheta+3H\dot\vartheta=0.
\end{equation}
If $f(12H_0^2)={}-1$, then we have the equation
\begin{equation}
\ddot\vartheta-H_0\dot\vartheta=0,
\end{equation}
which has the general solution:
\begin{equation}
\label{dSsolequ2} \vartheta=C_3+C_4e^{H_0t}.
\end{equation}
Thus, from (\ref{dSsolequ1}) and (\ref{dSsolequ2}), we get that the de
Sitter solution corresponds to
\begin{equation}
\label{dS-sol}
\vartheta_{dS}=C_1e^{H_0t}-\frac{B}{4H_0^2},
\end{equation}
where $C_1$ is an arbitrary constant. At $\Lambda=0$,
\begin{equation}
\label{dS-sol0}
\vartheta_{dS_0}=C_1e^{H_0t}-M_P^2.
\end{equation}
It is easy to see that the Minkowski
solutions (at $f(12H_0^2)={}-1$) correspond to
\begin{equation}
\vartheta_{M}=C(t-t_0),
\end{equation}
where $C$ and $t_0$ are arbitrary constants.

At $\vartheta=0$, Eqs.~(\ref{equa00psi}) and (\ref{equajjpsi}) have
the following nontrivial ($\psi$ is not a constant) de Sitter and Minkowski solutions:
\begin{itemize}
\item $H=0$, \ $\Lambda=0$, \ $\psi(t)=\frac{1}{\tilde{g}(t-t_0)}$, \
$f(0)={}-1$, \  $f'_1(0)$ is an arbitrary number.
\item $H=H_0\neq 0$, \  $\Lambda = 3M_P^2H_0^2$, \    $\psi(t) =
\frac{H_0}{\pm\tilde{g}+H_0\exp(H_0(t-t_0))}$, \  $f(0)={}-1$, \
$f'(0)$ is an arbitrary number.
\end{itemize}
These solutions do not change the value of the cosmological constant.
Solutions, which corresponds to $\vartheta(t)\not\equiv 0$ are more interesting.
The following equation for $\psi$ arises
\begin{equation}
f^{\prime}(R_0)\left(\dot\psi^2-\tilde{g}^2\psi^4+2H\dot\psi\psi+H^2\psi^2\right)=
\frac{B}{4H^2}-C_1e^{Ht},
\end{equation}
which, for $C_1=0$, yields the following first order differential equation for $\psi$:
\begin{equation}
\dot\psi^2+2H_0\dot\psi\psi=\tilde{g}^2\psi^4-H_0^2\psi^2-\frac{B}{4H_0^2f^{\prime}(12H_0^2)}.
\label{equpsideSitter}
\end{equation}
The trivial cases $c=0$ and $H_0=0$ have been considered above. In the
general case, Eq.~(\ref{equpsideSitter}) does not satisfy the Fuchs
conditions and, therefore, its  solutions are multivalued functions (see,
for example~\cite{Golubev}). For $C_1=0$ the solution $\psi(t)$ can been found by quadratures, namely
\begin{equation}
-\int\limits_0^\psi
{}\frac{1}{H_0\tilde{\psi}\pm\sqrt{\tilde{g}^2\tilde{\psi}^4-c}}d\tilde{\psi}=
t-t_0, \qquad c\equiv \frac{B}{4H_0^2f^{\prime}(12H_0^2)}.
\end{equation}
For nonzero values of $C_1$ the solution $\psi(t)$ can be found numerically.

\section{Conclusion}

In this paper, we have studied cosmological models with an $SU(2)$ Yang--Mills field.
The model with a minimally coupled  Yang--Mills field, described by the action (\ref{a1}),
includes second- and fourth-order terms of the Yang--Mills field strength tensor.
The second-order term can play the role of radiation, whereas the fourth-order one
plays the role of the cosmological constant.

We have shown that the function $\Fc(Z)$ can be reconstructed provided the Hubble parameter is given.
In particular, we have demonstrate that de Sitter solutions exist only in the trivial case, namely when $\Fc(Z)$ is a constant.

In order to obtain genuine de Sitter solutions, we have considered models in which the Yang--Mills field has a nonminimal coupling with gravity.
We have explicitly shown that this model, described by the action in (\ref{eq:2.1}), has de Sitter solutions even in the absence of a cosmological constant term. The de Sitter solutions correspond to the Yang--Mills fields which satisfy Eq.~(\ref{equpsideSitter}).
This equation includes an arbitrary parameter. Depending on the value of this parameter, it has been shown that it can be easily solved in quadratures or, in the most general case, numerically.

\medskip

\noindent {\bf Acknowledgements.} S.Yu.V. is grateful to the organizers of the XV International Conference on Symmetry Methods in Physics (Dubna, Russia, July 12--16, 2011) for the possibility to
present the results of this work and for financial support. E.E. and S.D.O. are supported in
part by MICINN (Spain), projects FIS2006-02842 and FIS2010-15640, by
the CPAN Consolider Ingenio Project, and by AGAUR (Generalitat de
Ca\-ta\-lu\-nya), contract 2009SGR-994. A.J.L.R. acknowledges a JAE fellowship from CSIC. S.Yu.V. is supported in
part by the RFBR grant 11-01-00894, by the Russian Ministry of
Education and Science under grants NSh-4142.2010.2 and NSh-3920.2012.2, and by contract CPAN10-PD12 (ICE, Barcelona, Spain).

\section*{Appendix}

In the FLRW spatially flat space-time
\begin{equation}
\begin{split}
\Gamma_{ij}^0&=H a^2\delta_{ij},\quad\Gamma^i_{j0}=H\delta^i_j,\quad\Gamma^\mu_{00}=0,\quad\Gamma^0_{\mu 0}=0,\\
\Box&={}-\partial_0\partial_0-3H\partial_0+\frac{1}{a(t)^2}\left(\partial_1\partial_1
+\partial_2\partial_2+\partial_3\partial_3\right),
\end{split}
\end{equation}
\begin{equation}
R_{00} = {}-3\left( \dot{H} + H^2 \right),
\hspace{2.7mm}
R_{0i} = 0\,,
\hspace{2.7mm}
R_{ij} = \left( \dot{H} + 3H^2 \right) g_{ij}\,,
\hspace{2.7mm}
R=6\left( \dot{H} + 2H^2 \right).
\label{eq:2.15}
\end{equation}
Using the field $A^b$ in the form (\ref{ansatzA}), we get two independent equations.
The first one reads:
\begin{equation}
\label{equ00}
R_{00}-\frac{1}{2}Rg_{00}=3H^2=\frac{1}{M_P^2} \left(T^{(\mathrm{YM})}_{00}+\Lambda\right).
\end{equation}

For any twice-differentiable functions $f(R)$ and $\mathcal{W}(t)$, we get
\begin{equation*}
\Box \left[ f^{\prime}(R) \mathcal{W} \right]=-f'''(R)\dot
R^2\mathcal{W}-2f''(R)\ddot R\dot{\mathcal{W}}-f'(R)\ddot{\mathcal{W}}
 -3H\left[f^{\prime\prime}(R) \dot R\mathcal{W}+ f^{\prime}(R) \dot{\mathcal{W}}\right] ,
\end{equation*}
\begin{equation*}
 g_{00} \Box \left[ f^{\prime}(R) \mathcal{W} \right]
- \nabla_{0} \partial_{0} \left[ f^{\prime}(R) \mathcal{W}
\right]=3H\left[f^{\prime\prime}(R) \dot R\mathcal{W}+ f^{\prime}(R)
\dot{\mathcal{W}} \right].
\end{equation*}
Using
\begin{equation*}
F^b_{\beta 0}F^b_{\alpha 0}g^{\alpha\beta}=F^b_{i 0}F^b_{i 0}g^{ii}=3\frac{\dot\phi^2}{a^2}, \quad
g^{\alpha\beta} F_{0\beta}^{b} F_{0\alpha}^{b}
-\frac{1}{4} g_{00} \mathcal{F}=\frac{3}{2}\left(\frac{\dot\phi^2}{a^2}+\frac{\tilde{g}^2\phi^4}{a^4}\right),
\end{equation*}
we obtain that Eq.~(\ref{equ00}) is equivalent to
\begin{equation}
\begin{split}
&2M_P^2H^2=(1+f(R))\left(\frac{\dot\phi^2}{a^2}+\frac{\tilde{g}^2\phi^4}{a^4}\right)- 6\left( \dot{H} + H^2 \right)f^{\prime}(R)
\left(\frac{\tilde{g}^2\phi^4}{a^4}-\frac{\dot\phi^2}{a^2}\right)+\frac{2\Lambda}{3}+{}\\&{}
+ 6H\left( \dot
Rf^{\prime\prime}(R)\left(\frac{\tilde{g}^2\phi^4}{a^4}
-\frac{\dot\phi^2}{a^2}\right) + 2
f^{\prime}(R)\left(\frac{2\tilde{g}^2\phi^3(\dot \phi-H\phi)}{a^4}
-\frac{\dot \phi(\ddot\phi-H\dot\phi)}{a^2}\right) \right).
\end{split}
\label{equa00}
\end{equation}

The second equation reads:
\begin{equation}
R_{ii}-\frac{1}{2}Rg_{ii}={}-g_{ii}\left(2\dot
H+3H^2\right)=\frac{1}{M_P^2}
\left(T^{(\mathrm{YM})}_{ii}+T^{(\mathrm{4})}_{ii}\right).
\label{equii}
\end{equation}
To calculate $T^{(\mathrm{YM})}_{ii}$ we use the following formulae (no summation over $i$)
\begin{equation}
F^b_{\beta i}F^b_{\alpha i}g^{\alpha\beta}=F^b_{0 i}F^b_{0
i}g^{00}+F^b_{j i}F^b_{j
i}g^{jj}={}-\dot\phi^2+2\tilde{g}^2\frac{\phi^4}{a^2},
\end{equation}
\begin{equation}
g^{\alpha\beta} F_{i\beta}^{b} F_{i\alpha}^{b}
-\frac{1}{4} g_{ii} \mathcal{F}=\frac{1}{2}\left(\frac{\tilde{g}^2\phi^4}{a^2}+{\dot\phi^2}\right),
\end{equation}
and get (\ref{equii}) in the following form
\begin{equation}
\begin{split}
\label{equajj}
    &-2\dot H -3H^2=\frac{1}{2M_P^2}\left[(1+f(R))\left(\frac{\tilde{g}^2\phi^4}{a^4}+\frac{\dot\phi^2}{a^2}\right)
   -2\Lambda +{}\right.\\&\left.{}+6(\dot H+3H^2)f'(R)\left(\frac{\tilde{g}^2\phi^4}{a^4}
    -\frac{\dot\phi^2}{a^2}\right)
    -6[\partial_0\partial_0+2H\partial_0]
     \left(f^{\prime}(R) \left(\frac{\tilde{g}^2\phi^4}{a^4}-\frac{\dot\phi^2}{a^2}\right) \right) \right].
\end{split}
\end{equation}
Using (\ref{psiphi}), we rewrite Eqs.~(\ref{equa00}) and (\ref{equajj}) in terms of $\psi(t)$, to get  (\ref{equa00psi}) and (\ref{equajjpsi}).

\end{document}